# A differentially rotating disc in a high-mass protostellar system

M.R. Pestalozzi[1], M. Elitzur[2], and J.E. Conway[3]

[1] Dept of Physics, University of Gothenburg, S-412 96, Göteborg, Sweden
e-mail: michele.pestalozzi@physics.gu.se
[2] Dept of Physics and Astronomy, Univ. of Kentucky, Lexington, KY 40506-0055, USA
e-mail: moshe@pa.uky.edu
[3] Onsala Space Observatory, S-439 92 Onsala, Sweden
e-mail: john.conway@chalmers.se

**ABSTRACT**

*Context.* A strong signature of a circumstellar disc around a high-mass protostar has been inferred from high resolution methanol maser observations in NGC 7538 -IRS1 N. This interpretation has however been challenged with a bipolar outflow proposed as an alternative explanation.
*Aims.* We compare the two proposed scenarios for best consistency with the observations.
*Methods.* Using a newly developed formalism we model the optical depth of the maser emission at each observed point in the map and LOS velocity for the two scenarios.
*Results.* We find that if the emission is symmetric around a central peak in both space and LOS velocity then it has to arise from an edge-on disc in sufficiently fast differential rotation. Disc models successfully fit ∼ 100 independent measurement points in position-velocity space with 4 free parameters to an overall accuracy of 3-4%. Solutions for Keplerian rotation require a central mass of at least 4 $M_\odot$. Close to best-fitting models are obtained if Keplerian motion is assumed around a central mass equaling ∼30 $M_\odot$ as inferred from other observations. In contrast we find that classical bipolar outflow models cannot fit the data, although could be applicable in other sources.
*Conclusions.* Our results strongly favour the differentially rotating disc hypothesis to describe the main feature of the 12.2 (and 6.7) GHz methanol maser emission in NGC 7538 IRS1 N. Furthermore, for Keplerian rotation around a ∼30,$M_\odot$ protostar we predict the position and velocity at which tangentially amplified masers should be detected in high dynamic range observations. Also, our model predicts the amplitude of the proper motion of some of the maser features in our data. Confirmation of a large central mass would strongly support the idea that even the highest mass stars (>20 $M_\odot$) form via accretion discs, similar to low-mass stars. Finally we note that our new formalism can readily be used to distinguish between discs and outflows for thermal emitting line sources as well as masers.

## 1. Introduction

### 1.1. High-mass star formation through accretion discs

While it is generally accepted that low-mass stars (M < 8 $M_\odot$) form via accretion discs the situation is less clear for more massive stars (see e.g. Zinnecker & Yorke 2007 for a review). High-mass protostars produce large radiation pressures which are able to reverse accretion flows and prevent the growth in mass of the central object. The simplest models suggest that stars more massive than 8 $M_\odot$ cannot form by standard accretion. To explain the observation of more massive stars alternative mechanisms have to be invoked such as competitive accretion, mergers of lower mass stars (see e.g. Bonnell et al. 2004) or accretion through a "trapped" hypercompact H II region (see e.g. Keto 2003 and references therein). It is therefore of great astrophysical interest to determine observationally whether high-mass protostars have accretion discs or not.

In recent years a number of claims of compact discs (< 1000 AU in radius) surrounding high-mass stars have been made (see the review of Cesaroni et al. 2007). These are based both on radio and millimetre continuum emission (see e.g. G192.16-3.82, Shepherd et al. 2001, GL490 Schreyer et al. 2006 and Orion-I Reid et al. 2007) as well as millimetre spectral line emission (IRAS 20216+4104 Cesaroni et al. 1997; Cesaroni et al. 2005, and Cep A, Patel et al. 2005). Maser emission observations include the equatorial disc outflow in Orion KL (Greenhill et al. 2004; Greenhill et al. 1998) and OH masers in IRAS 20216+4104 (Edris et al. 2005). Despite these results no disc has yet been observed around a protostar with mass larger than 20 $M_\odot$ (Cesaroni et al. 2007). Detailed numerical simulations invoking non-spherical accretion via dense discs are now able to model the formation of stars up to this mass but it is still an open question whether disc accretion is a valid formation mode for more massive stars (Zinnecker & Yorke 2007).

### 1.2. Methanol masers, discs and outflows

Methanol maser emission has been recognised as one of the best tracers of high-mass star formation regions and has been extensively searched for in the last two decades across the Milky Way. The most recent of these searches, after having covered 60% of the galactic plane, has yielded some 800 sources (see e.g. Green et al. 2009). Of these 520 objects were already reported in the literature and show a variety of characteristics (Pestalozzi et al.

*Send offprint requests to*: michele.pestalozzi@gmail.com

2005). All the known methanol masers are associated with high-mass star forming regions, most of them in very early stages of evolution, prior to the creation of an Ultra Compact (UC) H II region (see e.g. Ellingsen 1996; Walsh et al. 1998). A considerable number of these methanol masers are rich in spectral features that very often align both in space and in line-of-sight (LOS) velocity (Norris et al. 1998). These lines were interpreted as coming from edge-on rotating discs since in such systems the conditions for building up long velocity coherent amplification paths are naturally met.

Although the disc interpretation was the first model suggested for lines of methanol masers, other models have subsequently been proposed. It has been argued for instance that they arise in jets or outflows. In W3(OH) for instance, Moscadelli et al. (2002) model the methanol maser line as tracing the surface of a bipolar cone spiraling with a constant velocity around the cone axis. Another model proposes that the linear 6.7 GHz methanol masers are produced by planar shocks propagating nearly perpendicular to the line of sight in a rotating cloud (Dodson et al. 2004). Attempts have been made to distinguish between disc and outflow models by comparing the directions of the lines of masers and the $H_2$ outflow axis in sources where both are detected (De Buizer 2003). Among the sources that produce clear signature, in 6 the two axes are parallel, indicating an outflow origin for the masers, in 2 they are perpendicular, in agreement with the disc interpretation. Six additional sources are classified as "likely", though not definitively, parallel. While these results strengthen the outflow hypothesis, they indicate a disc origin in some cases.

### 1.3. Previous observations and interpretation of NGC 7538 IRS1 N methanol maser emission

Methanol maser emission in the high-mass star forming region NGC 7538 at 6.7 and 12.2 GHz was discovered by Menten (1991) and Batrla et al. (1987) respectively. Both masers were found to be coincident with the IR source NGC 7538 IRS1 N. Subsequently a number of high spatial resolution observations have been performed using the European VLBI Network (EVN[1]) and the Very Long Baseline Array (VLBA[2]). From these observations very accurate maps of the spatial and dynamical structure of the main spectral maser feature have been made. Recent MERLIN observations at lower spatial resolution have revealed that two other weak maser spectral features are associated with objects ≈1.5 arcminutes to the south of the main feature (close to IR sources IRS 9 and NGC 7538 -S). These results suggest that several massive stars are forming in the NGC 7538 region (Pestalozzi et al. 2006).

The main maser emission feature in NGC 7538 IRS 1 (feature 'A', see Pestalozzi et al. 2006 for the nomenclature) is ≈ 2 km s$^{-1}$ wide and has a peak flux of ≈350 Jy at 6.7 and 120 Jy at 12.2 GHz. It is seen projected on a UC H II region with brightness temperature of 10 000-15 000 K (Campbell 1984; Gaume et al. 1995). The source powering the UC H II region appears to be an O7 star (Akabane & Kuno 2005). Minier et al. (1998) recognised in the very linear shape both in space and LOS velocity of this spectral feature at both 6.7 and 12.2 GHz the first hint for a rotating disc seen edge-on. Assuming the *maser spots* to lie on the outer radius of a thin disc around a central star of 10 M$_\odot$, the authors suggested the radius at which the masers occurred to be several hundred AU. The same data was modelled in detail by Pestalozzi et al. (2004) assuming for the first time masing methanol over a range of radii (350AU - 1000AU), distributed in a disc in Keplerian rotation around a 30 M$_\odot$ star[3].

Challenging the above disc model mid-IR observations have been interpreted as showing dust emission at different temperatures tracing cavities excavated by an outflow (De Buizer & Minier 2005). Since these cavities are roughly oriented parallel to the line of methanol masers the maser emission has been interpreted as arising within the cavities of a collimated jet/outflow rather than a disc. A controversy is then in place: the methanol maser emission toward NGC 7538 IRS 1 seems to be validly interpreted as arising from a rotating disc seen edge-on or from a bipolar outflow, depending on the data considered for the analysis.

In even more recent work the outflow hypothesis for the methanol maser has been weakened. Kraus et al. (2006) find that the large scale CO outflow originating in NGC 7538 IRS 1 is probably precessing. Detailed precession models can be made consistent with a disc origin for the methanol masers and the large scale outflows with cavities traced by CO and IR emission.

In this paper we further develop the formalism used in Pestalozzi et al. (2004) with the aim to definitely test the competing disc/outflow explanations of the methanol masers in this object. We show that the bipolar outflow geometry is not able to reproduce the emission shown in the data. The only way to obtain a symmetric emission around a maximum both in space and LOS velocity is to model the maser as emitted by a differentially rotating disc seen edge-on.

## 2. Methanol maser data in NGC 7538 IRS1 N

The data modelled in this paper are shown in Figs. 1 and 2, where the maser optical depth is presented. This is obtained from the flux density maps using eq. 1 below. The 6.7 GHz data were taken in February 2001 using the EVN, the 12.2 GHz in March 2005 using the VLBA. The resolution in the data is indicated by the size of the beam in the lower right corner of the figures and is about 5 and 2 mas for the 6.7 and 12.2 GHz data respectively. The velocity resolution is 0.088 and 0.048 km s$^{-1}$ for the 6.7 and 12.2 GHz data respectively. Both the velocity integrated maps (Fig. 1) and the position-velocity $p, v$-diagrams (Fig. 2) contain a continuous smooth central feature extending over some 40 mas in RA. At the resolution of our VLBI observations this translates into ~8 to ~20 independent measurements across the central maser feature at 6.7 and 12.2 GHz, respectively. Such a large amount of data allows our models to be highly constrained. At displacements greater than 20 mas some weak disconnected regions of emission are visible. These outliers were already seen in the early 6.7 GHz interferometry data but were detected at 12.2 GHz only in the most recent observations.

From phase-reference astrometry it is seen that 6.7 and 12.2 GHz maser emissions are perfectly cospatial within 2 mas (see Pestalozzi et al. 2004). The cospatiality of the two emissions is also supported by the strong similarity between the maps in the first and third panels of Fig. 1. The former shows the 6.7 GHz emission and the latter the 12.2 GHz data convolved to the resolution of the 6.7 GHz map. This similarity is remarkable given that that the two maser lines are emitted by two different isomers of the methanol molecule, and hence are expected to trace different regions.

---

[1] The EVN is a joint facility of European, Chinese, South African and other radio astronomy institutes.

[2] The VLBA is operated from the National Radio Astronomy Observatory's Array Operation Centre in Socorro, NM.

[3] Note that in Pestalozzi et al. (2004) the value of 30 M$_\odot$ was adopted for consistency with observations at other wavelengths and was not obtained from the model.

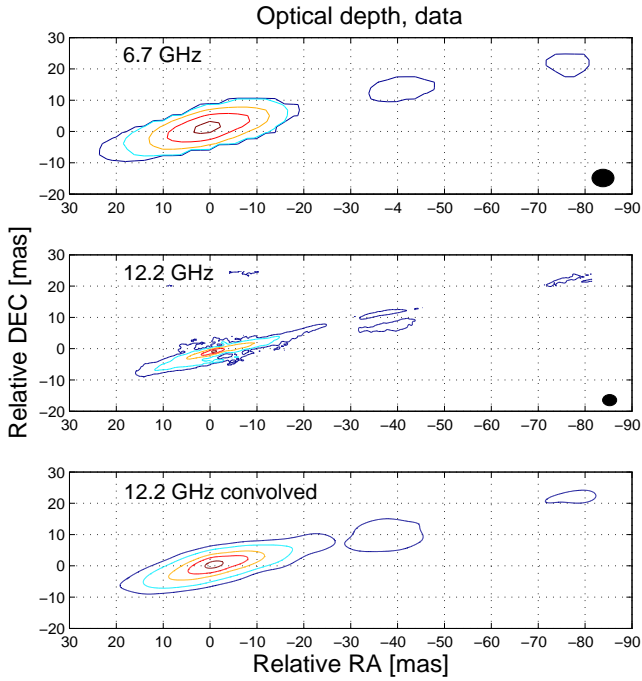

**Fig. 1.** Map of the velocity integrated optical depth $\tau$ of the maser emission in NGC 7538, obtained from eq. 1. From top to bottom: 6.7 GHz data, 12.2 GHz data, 12.2 GHz data convolved at the resolution of the 6.7 GHz data. The peak value of $\tau$ is 18 and 16 for the 6.7 and 12.2 GHz respectively, on the outliers it reaches 11 and 14. The beam size is indicated in the lower right corner of each panel. Contours are at 70,80,90,95,99% of the peak $\tau$ in all panels. Note that the lowest contour level is set by the dynamic range limit of our observations.

The superposition of the maser emission on bright centimetre continuum emission, together with the cospatiality of maser emission at two frequencies leads to the assumption that the maser is the result of the amplification of a background continuum. In that case the brightness at displacement $(x, y)$ from the centre of each frequency plane with LOS velocity $v$ is expressed as[4]:

$$I(x, y, v) = I_B \, e^{\tau(x,y,v)} \qquad (1)$$

where $I_B$ is the background continuum and $\tau$ the (negative) maser optical depth. This expression describes foreground amplification by any maser, irrespective of its degree of saturation (e.g., Elitzur 1992). The optical depth $\tau$ is therefore defined at every point in the cube and, knowing the brightness temperature of the background, it is a measured quantity, $\tau(x, y, v) = \ln[I(x, y, v)/I_B]$. Considering $I_B \approx 10^4$ K and the maser peak brightness per unit area $I(0, 0, 0)$ at the centre we have that $\tau(0, 0, 0) \equiv \tau_0 \approx 16$ and 18 for the 12.2 and 6.7 GHz respectively. In the outliers we have $\tau \approx 14$ and 11. This drop in optical depth between centre peak and outliers ($\approx 30$%) corresponds to a dynamic range in brightness of 100:1. Note that the shape of the contours in Figs.1 and 2 does not depend on the value of the background radiation $I_B$. The background brightness is only needed to obtain the value of the maser optical depth at the peak.

---
[4] For the model fitting to the radio interferometry data we adopt the radio convention to convert frequencies into LOS velocities: $f_{sky} = f_{rest}(1 - v_{los}/c)$, where $f$ is frequency

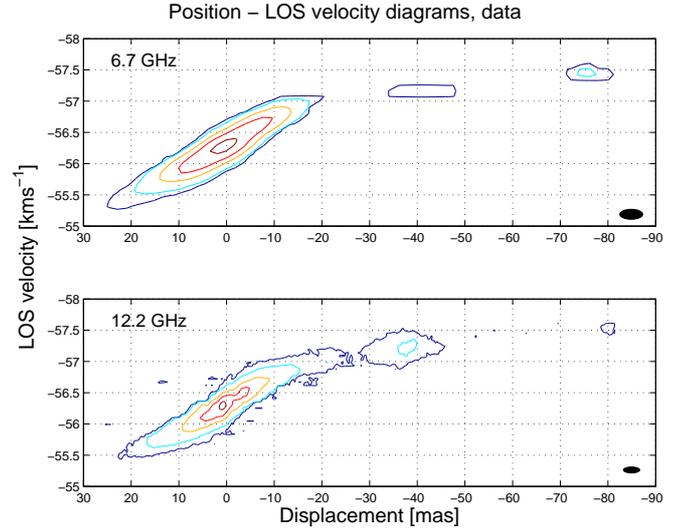

**Fig. 2.** LOS velocity versus spatial displacement ($p, v$–diagram) of $\tau$ of the main maser emission feature in NGC 7538. The top panel shows 6.7 GHz data, the bottom panel the 12.2 GHz data. The peak value of $\tau$ is 18 and 16 for the 6.7 and 12.2 GHz respectively, on the outliers features at 40 mas and 80 mas and it reaches 11 and 14. The beam size is indicated in the lower right corner of each panel. Contours are at 70,80,90,95,99% of the peak $\tau$ in both panels. As for Fig.1, the lowest contour level is set by the dynamic range limit of our observations.

There are two further important characteristics of these data. The first is that the bulk of the maser emission ($\pm 20$ mas) is symmetric about a maximum of emission both in space (Fig. 1) and in LOS velocity (Fig. 2). The second is that in the $p, v$–diagrams the emission is seen to bend away from the overall gradient at a displacement close to $\pm 10$-15 mas. These two facts together will be shown to be naturally explained by emission from a differentially rotating disc seen edge-on, whilst they are virtually impossible to reproduce with an outflow geometry. In the following discussion we will concentrate on the central part of the 12.2 GHz data ($\pm 20$ mas), as these show the highest spatial resolution.

## 3. The formalism: fundamental expression

Assuming that the observed maser emission is generated from amplified continuum radiation then the observed radiation is completely characterised by the maser amplification at each position (see eq. 1). Introduce Cartesian axes with $z$ toward the observer and denote by $\kappa(x, y, z)$ the (negative of the) linecenter maser absorption coefficient at an arbitrary position in the source and by $\phi(v)$ the local line profile. The maser optical depth is then $\tau(x, y, v) = \int \kappa(x, y, z)\phi(v - v_z)dz$, where $v_z$ is the LOS component of the local bulk velocity at $(x, y, z)$. Introduce $\tau_0 \equiv \tau(x_0, y_0, 0)$, the line-center optical depth at some conveniently chosen fiducial point $(x_0, y_0)$ in the plane of the sky. Then the maser optical depth can be written $\tau = \tau_0 T(x, y, v)$, where

$$T(x, y, v) = \int \eta(x, y, z) \, \phi(v - v_z) \, dz \qquad (2)$$

and where $\eta = \kappa/\tau_0$ contains all the effects of temperature, density and inversion gradients within the masing cloud.

The dimensionless function $T$, which is normalised such that $T(x_0, y_0, 0) = 1$, contains all the spatial and velocity information in the maser data cube. Because the data in Fig. 1 show a remarkably high aspect ratio we assume that the masers arise from a *flat* structure. We therefore rotated the cubes by $-17°$ and integrated them along the *height* axis to obtain displacement-LOS velocity diagrams ($p, v$–diagram, see Fig. 2). Hence in what follows we concentrate on modelling the function $T(x, v)$. In the cases of interest here, the spatial amplification profile $\eta$ is primarily a function of $r = \sqrt{x^2 + z^2}$, the radial distance from the center. The size scale of the system can be specified by some characteristic length $r_1$, the radial coordinate at some convenient point in the source; that is, the $\eta$-profile is a mathematical function of the scaled dimensionless variable $\rho = r/r_1$. Taking for $\phi$ a Gaussian profile with a width $\Delta v_D$ finally yields

$$T(x, v) = \int_{x/r_1}^{\infty} \exp\left[-\frac{1}{2}\left(\frac{v - v_z(x, \rho)}{\Delta v_D}\right)^2\right] \frac{\eta(\rho)\, d\rho}{\sqrt{1 - \left(\frac{x}{\rho\, r_1}\right)^2}}. \quad (3)$$

This general expression describes the optical depth of any maser, whether saturated or not. Saturation implies that the population inversion is affected by the propagating maser radiation, becoming one of the factors that shape the radial profile $\eta$. Here this is a moot issue because we limit ourselves to a parametric description of the $\eta$-profile without attempting to uncover its physical foundation. Saturation also causes broadening of the frequency profile of the maser absorption coefficient. This typically requires $\tau \gtrsim 15$ across the entire maser, and the profile broadening is then proportional to length in excess of this threshold (Elitzur 1992). With an optical depth of $\sim$ 16-18, saturation can be expected to have some effect here, but only at the very outer segments of the source where a decrease in temperature could have an opposite, offsetting effect. The width $\Delta v_D$ is therefore taken as constant, an approximation that is not expected to affect our fundamental conclusions.

While $\eta(\rho)$ describes the spatial distribution of the maser amplification, $v_z(x, \rho)$ contains the details of the dynamics of the system. Maser emission requires velocity coherence along the LOS—in order to participate in the maser action at a certain frequency $v$, the difference in LOS velocity of two points along a given direction $x$ cannot exceed the thermal width $\Delta v_D$. That is, once the geometry and dynamics of the system are defined, the variation of $v_z$ along any LOS determines the *coherence length* that controls the maser optical depth for that $x$. Figure 2 shows that the maser emission peaks at the center of the $p, v$–diagram. The challenge is to identify distributions $\eta(\rho)$ and $v_z(x, \rho)$ that reproduce this property.

In the next section we consider the application of the above formalism to edge-on discs. In Section 5 we fit the disc model to the NGC 7538 data and derive the best fitting disc parameters. In Section 6 we apply the formalism for the case of outflow geometry, and show why this model cannot fit the data shown in Figs.1 and 2.

## 4. Edge-on Disc Models

In an edge-on rotating disc, the LOS velocity $v_z$ and rotational velocity $v_{rot}$ of a point at distance $\rho$ from the center and azimuthal angle $\varphi$ (see Fig. 3) are related according to:

$$v_z(x, \rho) = v_{rot} \sin\varphi = v_{rot}(\rho)\frac{x}{\rho\, r_1} = \Omega(\rho)\, x, \quad (4)$$

where $\Omega(\rho)$ is the local angular velocity. Significantly, the LOS velocity is determined by the *angular* velocity, not the rotational velocity itself. Denote by $v_1$ the rotational velocity at $r_1$ and by $\Omega_1 = v_1/r_1$ the corresponding angular velocity. A fast rotating large disc and slowly rotating compact one will have widely different values of $v_1$ and $r_1$ yet they will produce the same LOS velocity field if they have the same $\Omega_1$ and radial variation of $\Omega$.

In this section we show from some simple considerations how the morphology of the data in Figs. 1 and 2 constrains the disc properties. In particular we demonstrate how the simultaneous appearance of a maximum of emission in the centre of both the map *and* the $p, v$–diagram can naturally be reproduced only by sufficiently fast differential rotation.

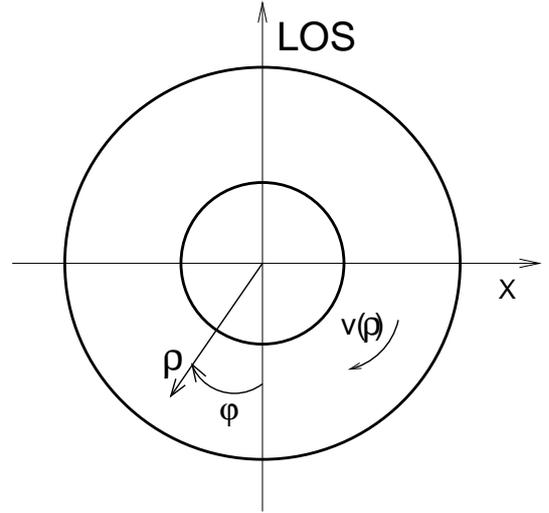

**Fig. 3.** Geometry for the disc model, with $v(\rho)$ the rotation velocity. The observer is in the $x - LOS$ plane.

### 4.1. The no-rotation case

In a non-rotating disc $v_z = 0$. Therefore the maser molecules are velocity-coherent along the entire source for every LOS, and the maser optical depth obeys $\tau(x, v) = \tau(x)\phi(v)$, where $\tau(x)$ is the line-center optical depth at position $x$. The positional variation of the amplification is controlled solely by the length of the amplifying column because the velocity profile $\phi$ is the same at every position. The $p, v$–diagram of the amplification contours in this case is shown in the top panel of Fig. 4. The contours show two distinct peaks, displaced symmetrically from the center at the two inner tangents, while *the center is a local minimum*. In any non-rotating disc with an inner radius in fact, the central LOS has a lower opacity than the LOS that is tangent to the inner radius, and this holds no matter what the $\eta(\rho)$ distribution is. This happens because, just like the LOS to the centre, the LOS tangent to the inner radius sample all $\rho$, but each interval of radius $d\rho$ corresponds to a longer path along the LOS than for the path toward the centre. Non-rotating discs will therefore never produce the central peak of the $p, v$–diagram evident in the data shown in Fig. 2.

### 4.2. Solid-body rotation

In solid-body rotation the angular velocity $\Omega$ does not depend on radius, therefore $v_z$ is constant along every LOS (eq. 4). Similar to the no-rotation case, the material remains fully velocity coherent along each LOS, only the velocity profile is now centered on $v = \Omega x$ instead of $v = 0$ so that the optical depth obeys $\tau(x,v) = \tau(x)\phi(v - \Omega x)$. As a result, the structure of amplification contours in the $p,v$–diagram remains the same, only rotated by the angle $\tan^{-1} \Omega$ (and slightly stretched to maintain the peak positions at the two tangents to the inner radius), as shown in the two bottom panels of Fig. 4. The center, $x = 0$, remains a local minimum, in complete analogy with the non-rotating disc. Solid-body rotation, too, will never produce the central peak observed in the $p,v$–diagram diagram.

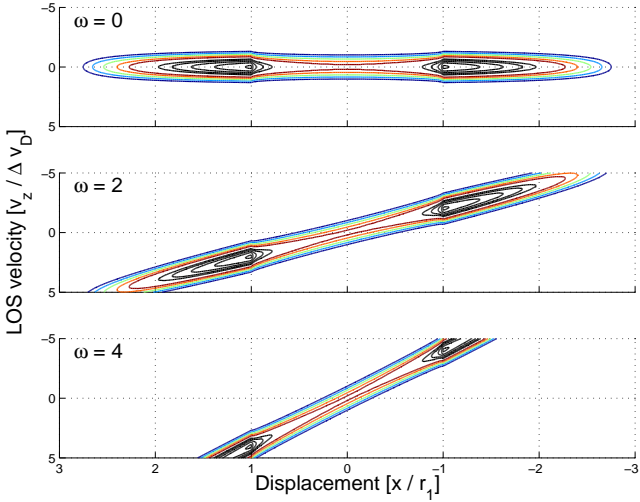

**Fig. 4.** $p,v$–diagram of maser optical depth (eq. 2) for edge-on discs with constant amplification ($\eta = const.$) between inner radius $r_1$ and outer radius $3r_1$. Contour values are normalised to the value at the diagram center and are spaced by 0.1, with the first contour at 0.6. Black contours indicate values greater than the central value. *Top:* A non-rotating disc. The emission maxima are along the tangents to the inner edges $x/r_1 = \pm 1$, the longest paths through the disc. *Mid and Bottom:* Same configuration as top panel only the disc rotates as a solid-body with the angular velocity $\omega \Delta v_D$, as indicated. Notice that all the maps are qualitatively identical except for the tilt, i.e. no-rotation and solid body rotation are two aspects of the same case.

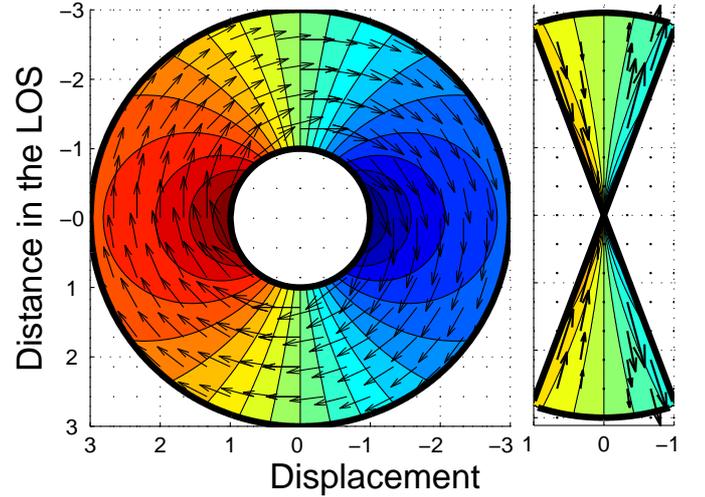

**Fig. 5.** Contours of constant LOS velocity ($v_z$) and the velocity fields that produce them. The $v_z$ contours are spaced by 0.1 of the highest value and are also colour-coded to indicate the Doppler velocity. The velocity vector field is indicated by arrows. *Left:* Keplerian rotating disc with masing material confined within inner and outer radii with ratio 1:3. *Right:* A bipolar flow that produces the same $p,v$–diagram as the disc geometry (see eq. 16 below), which has outflowing gas on the right side of the central LOS and inflowing gas on the left side as indicated by the arrows. The radial velocity was multiplied by a factor 2.5 for clarity. The masing material has been confined within a cone of opening angle 20°.

### 4.3. Differential rotation

The data shown in Fig. 2 cannot be reproduced by either a non-rotating disc or one rotating as a solid-body. Both scenarios fail because the disc geometry implies that the LOS through the disc center is a local minimum of pathlength. The introduction of differential rotation changes the situation fundamentally. Although the $x = 0$ amplification path is geometrically short, it maintains its full velocity coherence across the entire disc irrespective of the rotation law because the motion is perpendicular to the LOS and hence $v_z = 0$. In contrast, along any other path, the rotation velocity has a finite LOS component, with the consequence that segments of the path can get out of velocity coherence when $\Omega$ varies with radius. In this situation the optical depth at any LOS will no longer depend exclusively on the geometrical pathlength across the disc but will be limited by velocity coherence. This situation is represented in the left panel of Fig. 5, where coloured areas illustrate how the length of coherence paths changes with displacement in Keplerian differential rotation.

Differential rotation alone will not produce a maximum in the centre of the $p,v$–diagram if it is not sufficiently different from the two previously discussed cases. In general, differential rotation that is either too slow or to weakly dependent on $\rho$ will produce a twin-peak $p,v$–diagram, similar to the no-rotation or solid-body cases. For a given differential rotation law, a transition from a two-peaked to a single, central peaked $p,v$–diagram is expected only when the rotation velocity increases above a certain threshold. This is illustrated in Fig. 6, which shows the evolution of the $p,v$–diagram with increasing rotation velocity for a Keplerian disc with constant $\eta$. In this particular case, the threshold rotation velocity on the inner radius lies roughly between 4 and $6\Delta v_D$.

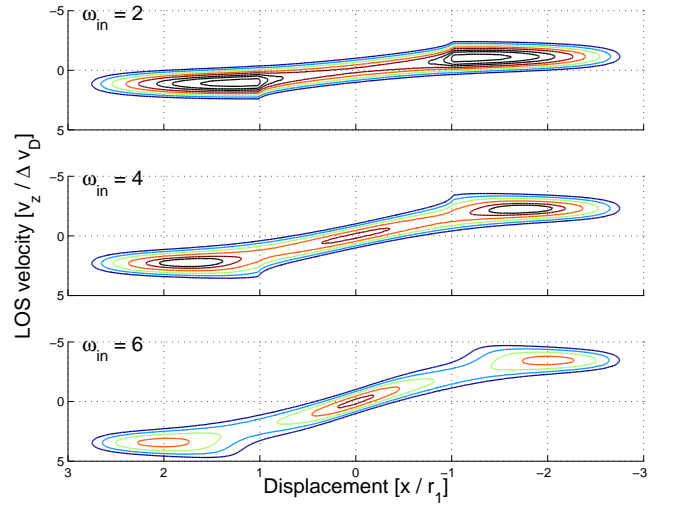

**Fig. 6.** Same as Fig. 4, only the disc is in Keplerian rotation, with angular velocity $\omega_{in} \Delta v_D$ on the inner radius. Notice that a maximum at the centre is obtained when the rotation is *fast enough* (bottom panel).

*4.4. The $p, v$–diagram "spine"*

In each of the displayed $p, v$–diagrams, the locus of points with strongest amplification at every LOS stands out as a clearly visible feature. We term this feature the *spine* of the diagram. In the case of solid-body rotation it is the straight line $v_z = \Omega x$, reverting to $v_z = 0$ for the non-rotating disc. Differential rotation introduces a curvature in the spine, evident in the varying orientation of highest-amplification contours in Fig. 6. Similar curvature is clearly visible also in the data in Fig. 2 shown by the change in velocity gradient at peak amplification between the inner and outer parts of the central maser feature. Spine analysis provides useful insight into the effect of the rotation on maser amplification. In particular, it provides an answer to the question: How fast must the differential rotation become for the central emission to turn from a local minimum to a local maximum?

The condition $\partial T(x, v)/\partial v = 0$ determines the relation between $v$ and $x$ on the spine, defining the spine curve $v_s(x)$ in the $p, v$–diagram. Maser amplification along this curve is the spine amplification $T_s = T(x, v_s)$. For a general spine analysis we express the rotation velocity as a power law, so that

$$v_{\rm rot} = v_1 \rho^\alpha, \quad \text{and} \quad v_z = v_1 \frac{x}{r_1} \rho^{\alpha-1} = \Omega_1 x \rho^{\alpha-1}. \quad (5)$$

Here $v_1$ is the rotation velocity at $r_1$ and $\Omega_1 = v_1/r_1$ is the angular velocity there. Keplerian rotation corresponds to $\alpha = -\frac{1}{2}$ while solid-body rotation is $\alpha = 1$. The spine condition then yields

$$v_s T_s = \Omega_1 \, x \int \exp\left[-\frac{1}{2}\left(\frac{v_s - v_z(\rho, x)}{\Delta v_{\rm D}}\right)^2\right] \frac{\rho^{\alpha-1} \eta(\rho) \, {\rm d}\rho}{\sqrt{1 - \left(\frac{x}{\rho r_1}\right)^2}} \quad (6)$$

The integral on the right differs from $T_s$ only by the additional term $\rho^{\alpha-1}$ in the integrand (cf eq. 3); when $\alpha = 1$ the two integrals are identical, and we recover the solid-body result $v_s = \Omega_1 x$ — the spine in this case is a linear structure tilted by angle $\tan^{-1} \Omega_1$. Consider now the centre of the $p, v$–diagram. At that point we always have $T(0, 0) = 1$, while the integral on the right of eq. 6 is $\langle \rho^{\alpha-1} \rangle$, where we introduced the notation $\langle \rho^p \rangle \equiv \int \rho^p \, \eta(\rho) {\rm d}\rho$ for the moments of the $\eta$-profile. Because angular velocity along the spine is $\Omega_s = v_s/x$, we have

$$\Omega_{s0} = \Omega_1 \, \langle \rho^{\alpha-1} \rangle \quad (7)$$

where $\Omega_{s0} \equiv \Omega_s(x = 0)$ is the slope of the spine at the origin of the $p, v$–diagram. This important result relates the measured quantity $\Omega_{s0}$ with the free parameters of the model. It is used extensively in our data analysis, as explained in section 5.

Equation 6 defines the spine only in an implicit form because $v_s$ enters on both sides. Straightforward series expansion in powers of $x/r_1$ produces in 2nd order

$$T_s(x) = 1 + \tfrac{1}{2} b \left(\frac{x}{r_1}\right)^2 + \ldots \quad (8)$$

where

$$b = \langle \rho^{-2} \rangle - \left(\frac{v_1}{\Delta v_{\rm D}}\right)^2 \left(\langle \rho^{2(\alpha-1)} \rangle - \langle \rho^{\alpha-1} \rangle^2\right)$$

This shows that the amplification along the spine will increase (decrease) with $x$ when $b$ is positive (negative), producing a local minimum (maximum) at the center of the $p, v$–diagram. A non-rotating disc has $v_1 = 0$, therefore $b = \langle \rho^{-2} \rangle > 0$ and the origin is always a local minimum in this case. Similarly, solid body rotation has $\alpha = 1$ so that again $b = \langle \rho^{-2} \rangle$, identical to the no-rotation case. A local maximum for $T_s$ at the origin requires $b < 0$, yielding

$$\left(\frac{v_1}{\Delta v_{\rm D}}\right)^2 > \frac{\langle \rho^{-2} \rangle}{\langle \rho^{2(\alpha-1)} \rangle - \langle \rho^{\alpha-1} \rangle^2} \quad (9)$$

This is the condition for a local peak at the origin of the $p, v$–diagram. It can be fulfilled only when both $\alpha \neq 1$, i.e., differential rotation, and $v_1$ exceeds $\Delta v_{\rm D}$ by some $\eta$-dependent prescribed factor, i.e., sufficiently fast rotation.

In addition to its potential to produce a central peak in the $p, v$–diagram, another fundamental property of differential rotation is the spine curvature. Similar to eq. 8, a 2nd order expansion for the angular velocity along the spine yields

$$\frac{v_s}{x} = \Omega_s(x) = \Omega_{s0} \left[1 - \tfrac{1}{2} c \left(\frac{x}{r_1}\right)^2 + \ldots\right] \quad (10)$$

where

$$c = \left(\frac{v_1}{\Delta v_{\rm D}}\right) \left(\langle \rho^{\alpha-1} \rangle \langle \rho^{-2} \rangle - \langle \rho^{\alpha-3} \rangle\right) +$$

$$\left(\frac{v_1}{\Delta v_{\rm D}}\right)^3 \left(2\langle \rho^{\alpha-1} \rangle^3 - 3\langle \rho^{\alpha-1} \rangle \langle \rho^{2(\alpha-1)} \rangle + \langle \rho^{3(\alpha-1)} \rangle\right)$$

and where $\Omega_{s0}$ is given in eq. 7. Solid-body rotation ($\alpha = 1$) yields $c = 0$ as it should (the angular velocity is constant on the spine), but in the case of differential rotation $c \neq 0$. Then the angular velocity on the spine deviates markedly from its value at the origin, and the spine curvature becomes noticeable, when the quadratic term becomes significant and the velocity coherence prescription begins to play a role in the determination of the amplification path length. The presence of a clear curvature in our data provides yet another strong evidence for the differentially rotating disc hypothesis.

## 5. Fit to the data

In this section we describe the results of the multi-dimensional fitting of disc models to our data. We first identify the natural scales in our problem for space and velocity with $r_1$ and $\Delta v_{\rm D}$, respectively. In addition to these scales, the problem contains the two unknown functions $v_z$ and $\eta$. Depending on their parametrisation, these functions will add to the total number of free parameters to be fitted.

Two quantities can be derived directly from the data without detailed analysis of the full $p, v$–diagram. The first one is the velocity scale $\Delta v_{\rm D} = 0.432 \, {\rm km \, s}^{-1}$, obtained from Gaussian fitting to the data at $x = 0$. The other quantity is the spine slope at the origin of the $p, v$–diagram, $D \Omega_{s0} = 0.064 \, {\rm km \, s}^{-1} \, {\rm mas}^{-1}$, where $D$ (= 2.7 kpc; see Blitz et al. 1982; Moscadelli et al. 2009) is the distance to the NGC 7538 region. Through eq. 7, $\Omega_{s0}$ provides a constraint on the free parameters we use in the detailed modeling.

*5.1. The function $\eta(\rho)$*

From eqs. 8 and 10 it is evident that both the shape on the spine and the intensity on it depend on the dynamics (represented by the value of $\alpha$) and on low order moments of the function $\eta(\rho)$. It follows that the exact form of $\eta(\rho)$ will be unimportant in fitting the data, only its overall properties such as width and kurtosis (lob-sidedness) will likely matter. This fact together with the need to minimise the number of fitted parameters suggest that a two-parameter function for $\eta$ is a practical choice. Pestalozzi et al. (2004) used a power-law function between inner and outer radii obtaining good results. Although this function has only two parameters (ratio of radii and power-law exponent) it also has unphysical sharp edges at the inner and outer radii, which we wish to avoid in this work. After some experimentation we settled on a modified log-normal function of the form

$$\eta(\rho) = A \rho^\gamma \exp\left[-\left(\frac{1 + \ln \rho}{\Delta}\right)^p\right], \quad (11)$$

where $A$ is determined from the normalisation $\int \eta(\rho)d\rho = 1$. To avoid blow-up when $\rho \to 0$, $\gamma$ must be positive and the argument of the exponential must remain negative, therefore $p$ must be an even integer. Further, without loss of generality we choose for $r_1$ the location of the maximum of $\eta$, that is the maximum of $\eta$ is at $\rho = 1$. This yields the following relation between $p$, $\gamma$ and $\Delta$:

$$\Delta = \left(\frac{p}{\gamma}\right)^{1/p} \quad (12)$$

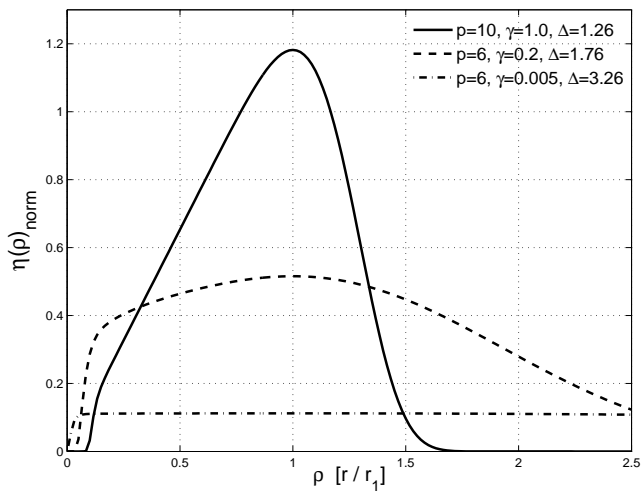

**Fig. 7.** Radial variation of the line-center absorption coefficient (eq. 11) for values of $p$ and $\gamma$ of three of the best-fit solutions in Table 1 with $\alpha = -0.5$ (solid), 0 (dashed) and 0.3 (dashed-dotted); the legend lists also the corresponding width $\Delta$ in each case (eq. 12). All profiles are normalized to unit area and have their maximum at $\rho = 1$ ($r = r_1$), though $r_1$ is different in each case (see Table 1).

This implies that the argument of the exponential term in eq. 11 becomes $-\gamma/p(1 + \ln \rho)^p$. The $\eta$-profile is then completely defined by the two parameters $p$ and $\gamma$. Figure 7 shows three examples for different combinations of $p$ and $\gamma$. While the function is continuous, there usually exist steeply rising "shoulders" of the profile. Large values of $p$ produce profiles with very steep shoulders while small values give smoothly decreasing functions. At a given $p$, the $\eta$-profile is very peaked around $\rho = 1$, similar to a thin ring, when $\gamma \gg 1$ (small $\Delta$; see eq. 12) and is very extended when $\gamma \approx 0$.

### 5.2. The function $v_z$ and dynamical mass

From eq. 5, the function for the LOS velocity $v_z$ adds two free parameters to the problem, $\Omega_1$ and $\alpha$, the angular velocity at $r_1$ and the index of the radial variation of $v_{\rm rot}$, respectively. The latter is set by the centripetal acceleration produced by the gravitational force of the mass distribution. Spherical distributions give a reasonable estimate of this dynamical mass even for the disc geometry because material inside and outside any $r$ has opposite effects in the two geometries that tend to cancel each other. The interior mass would produce a smaller centripetal acceleration when arranged as a sphere because some of the gravitational force goes into components perpendicular to the rotation plane. On the other hand, the mass outside $r$ would decrease the centripetal force by its outward pull when arranged in a disc but will have no effect when distributed in a spherical shell. The net effect is that at a given radius, a spherical mass produces rotational velocity similar to that of flat disc models having the same total mass (see e.g. Fig. 1 in Toomre 1963). Considering a spherical mass distribution and using eqs. 5 and 7, the dynamical mass inside radius $\rho$ is

$$M(\rho) = \frac{v_1^3}{G \Omega_1} \rho^{2\alpha+1}. \quad (13)$$

In Keplerian rotation $M$ does not depend on $\rho$, i.e., the cause of the dynamics can be considered a central point mass. Formally, the mass diverges when $\rho \to \infty$ for any $\alpha > -\frac{1}{2}$ but this divergence is meaningless in practice because the data probes only finite radii (although the mathematical expression for $\eta$ has no formal bounds). Differentiating the above expression for $M(\rho)$ shows that the mass density is proportional to $(2\alpha + 1)\rho^{2(\alpha-1)}$. Since the density must be positive, $\alpha$ must be larger than $-\frac{1}{2}$; Keplerian rotation provides the lower limit on $\alpha$.

### 5.3. Free parameters and fitting procedure

The masing absorption coefficient profile $\eta(\rho)$ is described by the two free parameters $p$ and $\gamma$, and the LOS velocity $v_z(x, \rho)$ by the two additional ones $\Omega_1$ and $\alpha$. Directly from the data, the spine tilt at the origin of the $p, v$–diagram is $D\Omega_{s0} = 0.064$ km s$^{-1}$ mas$^{-1}$, providing a constraint that reduces by one the number of free parameters: given $\alpha$, $p$ and $\gamma$, the angular velocity $\Omega_1$ is set by eq. 7. Two additional quantities are introduced by the fundamental expression (eq. 3) for the amplification map: the independent velocity scale $\Delta v_D$ and spatial scale $r_1$. The first is determined directly from the data, $\Delta v_D = 0.432$ km s$^{-1}$, while the second adds another free parameter, bringing their total number to four: $\alpha$, $r_1$, $p$ and $\gamma$. Here we carry out for the first time a fully unbiased search over these four parameters. This represents a significant improvement over Pestalozzi et al. (2004), where Keplerian rotation was assumed, a central mass of $30 M_\odot$ was taken from other observations and the best fit was found by optimising the two parameters defining $\eta$, which was chosen to be a single power-law between sharp cutoffs.

For a given set of the four free parameters, we determine the quality of the model by comparing the optical depth determined from eq. 3 with the data at every point in the central part of the $p, v$–diagram (within $\pm 20$ mas of the origin) where a measurement above the lowest detected opacity contour exists. Since spectral resolution is about 2 pixels in the velocity direction (corresponding to 0.1km/s) and FWHM beam size is 5 pixels or about 2 mas, there are 100 data measurements within the lowest contour.

As a quality estimator of each model we use the relative difference between model and data averaged over the detected part of the source:

$$X = \frac{1}{N} \sum_{i,j} \left| 1 - \frac{\tau(\theta_i, v_j)_{\rm model}}{\tau(\theta_i, v_j)_{\rm data}} \right|, \quad (14)$$

**Table 1.** Best-fit models of edge-on differentially rotating discs. The parameters $\Delta v_D = 0.432$ km s$^{-1}$ and $D\,\Omega_{s0} = 0.064$ km s$^{-1}$ mas$^{-1}$ are determined directly from the data and are common to all models. Assuming a distance $D = 2.7$ kpc, the free parameters $\alpha$, $p$, $\gamma$ and $r_1$ are determined from fits that minimize $X$, the error average over the entire central feature in the $p,v$–diagram (eq. 14). For each set of free parameters, eq. 7 determines $\Omega_1$, the angular velocity at $r_1$, and $v_1 = \Omega_1 r_1$ is the corresponding rotational velocity. $\rho_{\rm in}$ and $\rho_{\rm out}$ are the radii (in multiples of $r_1$) where the maser absorption coefficient drops to 5% of the maximum that it reaches at $r_1$. $M_{\rm in}$ and $M_{\rm out}$ are the corresponding enclosed dynamical masses (eq.13). The two are the same for Keplerian rotation, and the lower part tabulates additional solutions for this case with progressively increasing central mass.

| $\alpha$ | $p$ | $\gamma$ | $r_1$ [mas (AU)] | $X$ [%] | $\Omega_1$ [km s$^{-1}$ mas$^{-1}$] | $v_1$ [km s$^{-1}$] | $\rho_{\rm in}$, $\rho_{\rm out}$ | $M_{\rm in}$, $M_{\rm out}$ [$M_\odot$] |
|---|---|---|---|---|---|---|---|---|
| +0.7 | 4 | 0.005 | 0.2 (0.5) | 4.45 | 0.213 | 0.04 | $10^{-7}$, 409.3 | $1.4 \times 10^{-23}$, 1.7 |
| +0.6 | 4 | 0.007 | 0.4 (1.1) | 3.58 | 0.239 | 0.1 | $10^{-7}$, 233.4 | $5.1 \times 10^{-21}$, 2.1 |
| +0.5 | 4 | 0.009 | 0.8 (2.1) | 2.81 | 0.255 | 0.2 | $10^{-7}$, 158.1 | $9.6 \times 10^{-18}$, 2.4 |
| +0.4 | 4 | 0.01 | 1.1 (2.9) | 2.70 | 0.279 | 0.3 | $10^{-7}$, 135.3 | $7.4 \times 10^{-17}$, 2.0 |
| +0.3 | 6 | 0.005 | 14 (39) | 2.92 | 0.110 | 1.6 | $10^{-7}$, 18.5 | $7.2 \times 10^{-13}$, 12.1 |
| +0.2 | 6 | 0.01 | 18 (50) | 3.06 | 0.086 | 1.6 | $10^{-7}$, 12.1 | $2.3 \times 10^{-11}$, 4.8 |
| +0.1 | 6 | 0.05 | 32 (88) | 3.07 | 0.055 | 1.8 | $10^{-7}$, 5.4 | $1.3 \times 10^{-9}$, 2.4 |
| 0 | 6 | 0.2 | 45 (123) | 3.09 | 0.044 | 2.0 | $10^{-7}$, 3.2 | $5.5 \times 10^{-8}$, 1.8 |
| −0.1 | 6 | 0.3 | 49 (133) | 3.09 | 0.040 | 2.0 | $10^{-7}$, 2.8 | $1.5 \times 10^{-6}$, 1.4 |
| −0.2 | 6 | 0.5 | 57 (153) | 3.08 | 0.039 | 2.2 | 0.07, 2.4 | 0.2, 1.4 |
| −0.3 | 6 | 0.7 | 63 (171) | 3.07 | 0.038 | 2.4 | 0.07, 2.2 | 0.4, 1.5 |
| −0.4 | 8 | 0.9 | 90 (242) | 3.04 | 0.035 | 3.1 | 0.07, 1.7 | 1.5, 2.9 |
| −0.5 | 10 | 1.0 | 110 (298) | 2.99 | 0.032 | 3.5 | 0.07, 1.5 | 4.1 |
| −0.5 | 12 | 1.0 | 160 (431) | 3.04 | 0.031 | 4.9 | 0.07, 1.5 | 12 |
| −0.5 | 12 | 1.0 | 220 (594) | 3.07 | 0.031 | 6.8 | 0.07, 1.5 | 31 |
| −0.5 | 12 | 1.0 | 440 (1187) | 3.09 | 0.031 | 13.5 | 0.07, 1.5 | 245 |
| −0.5 | 12 | 1.0 | 1000 (2700) | 3.09 | 0.031 | 30.8 | 0.07, 1.5 | 2881 |

where $N$ is a normalisation parameter equalling the number of pixels in the fitted $p,v$–diagram; hence $X$ is a measure of the mean fractional error of the model opacity compared to that observed in the data. We chose this procedure after some experimentation. Although the measurement errors generally increase toward the edge of the central feature, all points were given equal weight in this averaging procedure because of the difficulty in estimating the noise in data determined from the logarithm of the intensity map. For this reason we also did not attempt a formal reduced $\chi^2$ analysis. The choice of the absolute value of the difference between model and data is in general preferable to the square of the difference, which tends to emphasize noisy points (see Press et al. 2007).

While the quality estimator $X$ includes only the central feature ($|x| \leq 20$ mas), larger displacements from the origin play an equally important role in constraining disc models. As discussed in §4.3, and displayed in Fig. 6, differential rotation that is not sufficiently fast will produce *tangential amplification features* at displacements $\gtrsim 5$ times larger than the central feature size, i.e., $|x| \gtrsim 100$ mas. Our data show that such features do not exist to within the sensitivity of the observations, whose dynamic range is ∼100:1. This implies that maser optical depth in the tangential features must be less than 70% of $\tau_0$, its value at the origin of the $p,v$–diagram. Acceptable models must meet this criterion in addition to producing a suitably small $X$.

### 5.4. Results

Table 1 presents a series of best fits. Starting from the different values of $\alpha$ listed in the first column we found the combination of $r_1$, $p$ and $\gamma$ that minimised $X$ (eq. 14) and produced acceptable models. The quality of all tabulated fits is high. Figure 8 shows the detailed comparison of model and data for the tabulated Keplerian model with 31 $M_\odot$. This model produces a formal error of $X = 3.07\%$, and the figure's right panel shows that it describes the data perfectly at the center of the $p,v$–diagram and to better than 10% for the large majority of calculated points within 20 mas of the origin. As is evident from the table, every rotation curve in the range $-0.5 \leq \alpha \leq 0.7$ is capable of producing a high-quality fit with an average error of only 3-4%. The corresponding amplification profiles $\eta(\rho)$ vary from broad distributions at large $\alpha$ to peaked shapes as $\alpha$ decreases toward Keplerian rotation. This is evident from the few examples plotted in Fig. 7 as well as the tabulated values of the radii $\rho_{\rm in}$ and $\rho_{\rm out}$ where $\eta$ decreases to 5% of the peak value. Moreover, the few entries listed for Keplerian rotation show that an unlimited number of models are producing essentially the same quality fits in this case. These different entries have nearly the same $\eta$ and $\Omega_1$, differing from each other only in the size scale $r_1$, which can be increased without bound. The same behaviour has been found for models with $\alpha \leq +0.4$. For those models Table 1 contains only the solutions with minimal $r_1$.

The reason for the degeneracy among so many highly successful models is that the data sample only a small portion of the full $p,v$–diagram generated by the disc. As an example, Fig. 9 shows the full map for the 31 $M_\odot$ Keplerian model, drawn to a level of 1% of peak value. While the map extends to $|x| > 300$ mas, the data cover only the inner $|x| \lesssim 20$ mas (red contours in the figure). This has a profound impact on the sensitivity of the model fitting. In fact, as is evident from the fundamental expression eq. 3, the dependence on the scale $r_1$ comes only from the integration lower limit and the curvature term in the denominator, and in both of them it enters as $x/r_1$. While the data extend only to 20 mas, $r_1$ is at least 110 mas for Keplerian rotation (and 220 mas for the model displayed in Fig. 9), so $x/r_1 \ll 1$ and $(x/r_1)^2$ is even smaller still at all measured points. Therefore, the entire data cube could be adequately described by a series expansion limited to the first few powers of $x/r_1$, similar to the spine analysis presented in §4.3. The expansion coefficients involve only moments of the amplification profile $\eta$, which can be easily adjusted with a suitable choice of its parameters; this is evident from the expressions for the spine (eqs. 8 and 10), the data most prominent feature. Furthermore, because $x/r_1$ is so small, all acceptable models are practically indistinguishable from the limit $x/r_1 \to 0$, which can be formally obtained by taking $r_1 \to \infty$ while holding $\Omega_1$ constant. This formal limit implies $v_1 \to \infty$, i.e., an infinite mass (see eq. 13). As can be seen from Table 1, every Keplerian entry with 12 $M_\odot$ and above has the same profile $\eta$ (same $p$ and $\gamma$) and the same $\Omega_1$, differing only in the scale size $r_1$; that is, the infinite mass limit is reached already at 12 $M_\odot$ in the case of Keplerian rotation. Similar behavior is found for the other tabulated values with $\alpha \le +0.4$. Although $r_1$ decreases as $\alpha$ increases, the profile $\eta$ becomes much more extended and the bulk of the integration originates from radii much larger than the observed displacements, yielding a similar behavior to the Keplerian case, which is concentrated around $\rho \sim 1$.

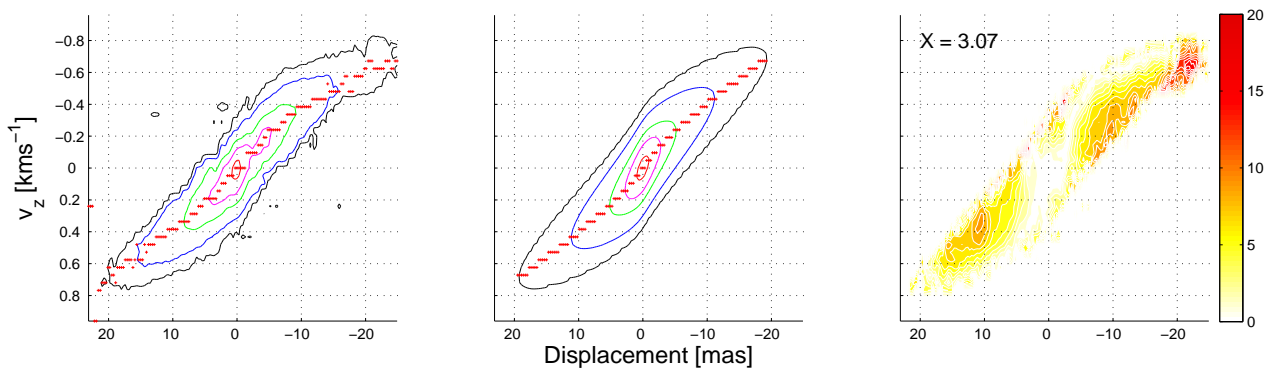

**Fig. 8.** Data (*left*) and modeling (*middle*) of the bulk of the 12.2 GHz maser emission (within ±20 mas of the feature centre). Contours are at 70, 80, 90, 95 and 99% of the peak optical depth ($\tau_0 = 16$). Red dots show the amplification spine (see §4.4). *Right*: Map of the model relative error at every position in the $p, v$–diagram, corresponding to the individual terms in the sum in eq. 14. Note the color bar on the right for the error magnitude in percent. The error averaged over the entire map is 3.07%. The model corresponds to an edge-on Keplerian disc around a 31$M_\odot$ star (see Table 1 for details).

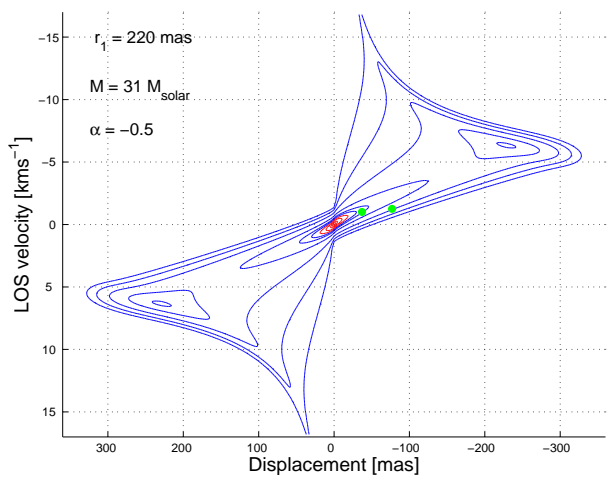

**Fig. 9.** Full map of the best fitting Keplerian model with a 31$M_\odot$ central protostar. Contours are plotted at 1, 3, 10, 30, 50, 60, 70, 80, 90, 95 and 99 % of the peak optical depth $\tau_0 = 16$. The red contours outline the region visible in our data (> 70% of $\tau_0$); this is the region shown in Fig. 8. Notice the small fraction of the full extent of the masing disc covered by the available data. The green dots indicate the positions of the outliers.

As this discussion shows, analysis of the available data is quite insensitive to the free parameter $r_1$, which is equivalent to $v_1$ (= $\Omega_1 r_1$). The only meaningful constraint on the latter is a lower limit on $v_1/\Delta v_D$ to ensure that the origin of the $p, v$–diagram is a local maximum (eq. 9) and that the tangential emission is sufficiently suppressed. Removal of the degeneracy among the best-fit models requires interferometry with a higher dynamic range, capable of detecting the tangential features. Current data do show emission at displacements between 30 mas and 80 mas (see figs. 1, 2 and 9), and we term these features "outliers." We do not think these outliers are the tangential features predicted by the models because their displacements are much smaller than the disc outer radius and they are not located quite on the extended spine, although close to it. Instead, the outliers are most likely regions of chance enhanced LOS coherence close to the spine, as suggested by their location and by the fact that they occur on only one side of the disc. Because of the maser exponential amplification, an enhancement of only 10% in the local value of $\tau$ would suffice to produce these outliers. Also, small deviations from axial symmetry (outliers only on one side) could be easily generated by, e.g., spiral density waves or a small warp that brings the East side of the maser disc slightly closer to edge-on than the West side, producing better LOS alignment of maser molecules (and hence outliers only on one side of the disc). Such a warp might be consistent with the disc precession in NGC 7538 IRS 1, inferred by Kraus et al. (2006).

### 5.5. Dynamical considerations

Edge-on discs in differential rotation fully capture the structure of maser amplification in the limited region of the $p, v$–diagram covered by our data. Most disc parameters are largely irrelevant as long as the rotation is sufficiently fast. This degeneracy makes it impossible to determine the disc properties purely from best-fit analysis of the maser data. We must invoke additional considerations to narrow down the range of acceptable disc parameters.

For each successful model, Table 1 lists the dynamical masses $M_{\rm in}$ and $M_{\rm out}$, calculated from eq. 13, contained within $\rho_{\rm in}$ and $\rho_{\rm out}$, the respective radii where $\eta$ decreases to 5% of its peak value on each side of the peak at $r_1$ ($\rho = 1$). These radii effectively mark the inner and outer boundaries of the disc maser region. As noted above, most sets of parameters allow $r_1$, and the corresponding $v_1$, to increase indefinitely; both $\rho_{\rm in}$ and $\rho_{\rm out}$ are left intact, though. As is evident from eq. 13, both $M_{\rm in}$ and $M_{\rm out}$ increase then without bound but their ratio remains constant. The tabulated $M_{\rm in}$ show that all models with $\alpha \geq -0.1$ are essentially devoid of any mass interior to the maser region. These models correspond to self-gravitating discs without any central object, either a star or just a central bulge, and thus can be discarded as unlikely to arise in realistic situations. Models with $-0.2 \geq \alpha \geq -0.4$ do contain a sizeable central object and thus are more likely to correspond to possible configurations. However, in each case the disc contains a significant fraction of the full mass, an inherently unstable situation: as shown by Adams et al. (1989), such discs are unstable to growth of eccentric distortions arising from small deviations between the positions of the star and the system centre of mass. Such unstable systems are unlikely to harbour the remarkably smooth, regular structure observed in the maser central feature. The problem is avoided only in the Keplerian models, where the mass in the disc is negligible in comparison with that in the central object.

Keplerian models are the only tabulated ones to offer stable physical systems, yielding a lower bound of $4.1\,\rm M_\odot$ on the central mass; all larger masses produce equally successful fits while smaller masses produce tangential features that conflict with the data. Therefore, we can conclude with reasonable confidence that *the maser central feature arises from an edge-on Keplerian disc around a central star heavier than* $4\,\rm M_\odot$. Kraus et al. (2006) conclude that the most plausible explanation for the jet precession around NGC 7538 IRS1 is a circumbinary disc around a binary pair separated by ~7 mas (~19 AU), which is much smaller than the minimal value of 110 mas for $r_1$. The maser disc fully encompasses the binary pair, whose total mass therefore must exceed $4\,\rm M_\odot$. This lower limit on the central mass is consistent with the ~$30\,\rm M_\odot$ inferred from the O7 spectral type of IRS1 (Akabane & Kuno 2005) but the degeneracy of our model fits prevents us from conclusively identifying this star with the center of the maser disc. However, the apparent alignment of the disc with the "waistline" of the associated ultra-compact H II region strongly suggests that this is indeed the case. The alternative would require a chance coincidence of this star with a foreground disc around a lower mass star, which seems less likely.

## 6. Bipolar outflow

In the previous two sections we considered the application of the general opacity formalism of eq. 3 to the specific case of an edge-on rotating disc. However, as noted in §1.3 it has been proposed that the linear maser feature in NGC 7538 IRS 1 is instead due to an outflow. This alternative can readily be modelled using the same general expression by modifying the velocity field. Consider a flow with radial velocity $v_{\rm out}$. The LOS velocity of a point at distance $\rho$ from the center and azimuthal angle $\varphi$ is now

$$v_z(x,\rho) = v_{\rm out} \cos\varphi = v_{\rm out}\frac{z}{\rho r_1} \qquad (15)$$

(see Fig. 10). Other than a different relation between velocity and its LOS component (cf. eq. 4), the outflow is handled exactly the same as the edge-on disc. Is it possible devise a radial configuration that produces the observed central maximum in the $p, v$–diagram? In this section we first demonstrate that, when measuring only Doppler velocity components, every disc model can be formally considered as if arising from radial motions, but this formal equivalence is unphysical as it requires alternating infall and outflow along adjacent radial rays (§6.1). We then show that physical outflow models cannot reproduce the observed central peak (§6.2).

### 6.1. Transferring disc solutions to radial flow models

Comparison of eqs. 4 and 15 shows that, given an edge-on disc rotating with velocity $v_{\rm rot}$, we can always define an equivalent radial flow with the velocity

$$v_{\rm out}^{\rm eq} = v_{\rm rot}\frac{x}{z} \qquad (16)$$

and the two will have the exact same LOS velocity field. As long as we measure only the LOS component of the velocity, it would be impossible to distinguish between the two configurations. The right panel of Fig. 5 shows the radial flow velocity field equivalent to the Keplerian disc shown in the figure's left panel. Since these two very different morphologies will produce the same LOS pattern, no direct observations can differentiate between them without measuring proper motions. However, although they will produce identical maps and $p, v$–diagrams, the physical foundations of the two configurations differ significantly. The Keplerian field arises naturally from rotation around a central star. In contrast, the equivalent radial motion does not arise from any plausible physical scenario, requiring zero velocity along the central LOS, outflow to one side of this direction and inflow to the other. This peculiar morphology is essential for reproducing the central peak and ordered gradients in the observed $p, v$–diagram. Therefore, although radial flow configurations can be mathematically devised to fit the data as well as edge-on rotating discs, they represent highly contrived velocity fields that are not produced by any physical model.

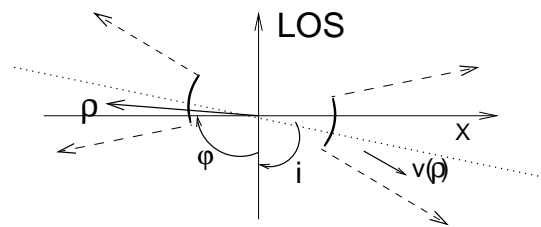

**Fig. 10.** Geometry of a physical outflow model whose axis is inclined at an angle $i$ to the LOS. Distance along the outflow is measured by $\rho$. The radius $\rho = \rho_1$ marks an injection point for material into the outflow (see text). Beyond this point the outflow has a position dependent bulk flow velocity $v(\rho)$ (dashed arrows) and mass flux conservation is assumed. In general the outflow has a cross-sectional radius $w(\rho)$, which for the conical flow illustrated is proportional to $\rho$.

### 6.2. A physical outflow scenario

Rather than starting from the disc solution as in §6.1 we can ab initio consider possible collimated outflows which could fit our data. Consider a narrow central symmetric outflow having cross-sectional radius $w$, density $n$ and velocity $v$, all only depending on the distance from the centre $\rho$ (see Fig. 10). If the inclination of the outflow axis to the LOS $i$ is significantly larger than the outflow opening angle then every LOS samples only one value of the LOS component of $v$, $v_z(\rho)$. From the $p, v$–diagram of such an outflow it is then possible to directly derive the outflow velocity function $v(\rho)$. Within this framework, our data in Fig. 2 show constant acceleration for $|x| \leq 10$ mas and then a reduction in acceleration at larger displacements, indicated by the "bend" of the overall gradient. This gives in general that $v(\rho) \propto \rho$. Mass flux conservation within the outflow implies $w^2 n v = const$ and the total gas column density across the jet varies as $w n \propto 1/(w v)$. Assuming that the maser optical depth $\tau$ is proportional to this column density, even without having any constraints on the function $w$ we can say that $\tau$ will decline at least as $1/\rho$. This dependence becomes even stronger if we further assume that the jet cross section $w$ is increasing with $\rho$ (e.g. conical outflow). The relationship between $\tau$ and position along the outflow can obviously not be extended to $\rho = 0$, and in any case any outflow must have a reservoir from which outflowing material is drawn. We can simulate such a reservoir with an injection point $\rho = \rho_1$, indicated by the thick lines in Fig. 10. If the reservoir is an accretion disc $\rho_1$ would correspond to the half of the disc thickness. At $\rho > \rho_1$ mass flow rate is conserved, inside $\rho_1$ this is no longer the case due to complex dynamics. This fact introduces a discontinuity in the outflow geometry that is not present when modelling an edge-on disc. This discontinuity is inherent in the nature of the bipolar outflow geometry and could therefore be detected in our observations, provided observations are done at a high enough spatial resolution. Our observations suggest that if such a discontinuity is present in our data this has to be on a scale that is smaller than the half of our resolving beam width, i.e. ~1 mas.

From our data we can express a strict constraint on the orientation of the outflow axis. This comes from making the (reasonable) assumption that the outflow velocity at the injection radius is approximately equal to the internal velocity dispersion $\Delta v_D$. The inclination angle is then obtained from the ratio of the LOS velocity at a certain $x$ and $\Delta v_D$. As noted above, the injection point lies at less than a beam FWHM projected distance from the centre (i.e. ~1 mas) and therefore has a LOS velocity of $0.064$ km s$^{-1}$ (see estimate of the overall gradient in the data). Using $\Delta v_D = 0.432$ km s$^{-1}$ we have that $i \geq \cos^{-1}(0.064/0.432) \approx 81°$, i.e. the outflow is almost in the plane of the sky.

Figure 11 shows an example $p, v$–diagram of a bipolar outflow illustrating the effect of a varying inclination $i$ to the LOS. The $1/\rho$-dependence makes $\tau$ decline very quickly with displacement, in strong conflict with our data, which show a smooth structure over about 10 beams radius. Denote by $\mathcal{R}$ the ratio in maser optical depth between one beam out at $x = 2$ and ten beams out, $x = 20$ mas. In a constant cross-section outflow, $\mathcal{R} \sim 5.5$ from the ratio of the corresponding LOS velocities. An increasing cross-section would make the ratio larger still—a conical outflow with apex on the central star would give $\mathcal{R} \approx 550$. In contrast, the observations show that $\mathcal{R} \approx 1.1$ since $\tau$ only decreases from 15.5 at one beam FWHM from the peak to 11 at ten beams out. It is impossible to reconcile such extended maser emission with the rapid density decline in outflows.

An important assumption in our model is that maser opacity scales with total gas column density. While the factors that affect maser opacity can be very complex we argue that the change in total gas column density in an outflow is so rapid (at least $1/\rho$) that this must strongly affect maser opacity. The only alternative to mitigate this effect is a fine-tuning of the maser opacity per total gas column density to almost exactly compensate for $1/\rho$

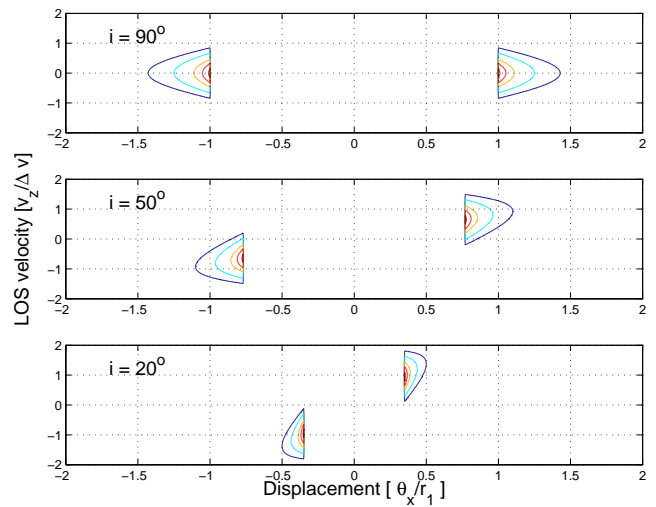

**Fig. 11.** $p, v$–diagram of maser optical depth for the bi-conical outflow sketched in Fig. 10. Each panel shows a different inclination $i$ of the outflow axis to the LOS. Contours are drawn at 70, 80, 90, 95 and 99% of the peak optical depth. The outflow has a constantly accelerating velocity, starting from $\Delta v_D$ at the injection radius $\rho_1$. Maser opacity beyond $\rho_1$ is proportional to total gas column density. The opacity and dynamics inside the injection radius are ill-defined: in this example the opacity there is taken as zero; quiescent maser emission from material within $\rho \leq \rho_1$ would add in each panel a horizontal stripe about the displacement axis, similar to the top panel of figure 4.

dependence of the total gas column density. This we consider highly unlikely.

A final possibility to consider is that a bipolar outflow could produce maser emission from interaction with ambient clumps of the ISM. In this case the fall off in gas column density within the outflow itself is not directly relevant, the important quantity is the gas column density of the ambient ISM. Such models will produce a string of maser features, symmetrically displaced around a central minimum in the $p, v$–diagram. Although numerous sources do display such structures and thus could be explained with bipolar outflows, these models cannot explain the smooth central peak observed in NGC 7538 IRS1 N.

## 7. Discussion and Conclusions

### 7.1. Discussion

The dominant methanol maser feature in NGC 7538 IRS 1 N is remarkable amongst observed maser structures for its smoothness and symmetry. At 12 GHz (see Fig. 1) the overall linear structure extends over 50 beams. The central feature extends over 20 beams and shows a clear 'S' shaped symmetry in its $p, v$–diagram. Such a structure must come from some coherent geometry which controls the gas flow; both discs and outflows have been proposed to provide this controlling geometry. In this paper we have applied a general formalism to simulate these two cases and we find that discs can readily fit the data. In contrast outflows predict gaps in the emissivity at the centre of the $p, v$–diagram and rapid fall-off of the maser opacity with projected distance, all contrary to our observations. Only extremely contrived outflow models are able to fit the data and therefore we rule these out.

The fundamental reason why discs are superior to outflows for fitting the observations is the different functional forms of the opacity versus projected distance which comes from their different symmetries. At the front of a disc the column density per velocity coherence length scales as $1 + kx^2$ (see §4.4 and eq. 8). For a sufficiently rapid differentially rotating disc $k$ can be negative in which case the centre becomes a local maximum in opacity, just as observed. The absolute value of $k$ however remains small enough compared to the range of $x$ so that the opacity can, as observed, change only by a few tens of percent for changes of a factor of 10 in $x$. In contrast for an outflow the opacity per unit velocity width scales as $x^{-n}$ down to some minimum $x$ where $n$ is at least 1 (see §6.2). This means that the change in opacity with projected distance is very fast and can only be reconciled with the data if one proposes a very contrived behavior for the ratio of maser opacity to total column density. An additional fundamental property of outflows is that there must be discontinuity in its properties at the minimum $x$; a discontinuity which we do not see.

We believe that our NGC 7538 data provide perhaps the best purely dynamical evidence to date for a rotating disc in a high-mass star forming region, from any spectral line observation, thermal or maser. Amongst maser sources the detection of a single smooth continuous symmetrical structure spanning 20 beams is extremely unusual as other maser sources in star-forming regions generally consist of a series of disconnected spots. Furthermore this number of beams across the source is much larger than for thermal emission observations: these usually span only a few beams just providing average velocity gradients. In contrast, in our $p, v$–diagram data we find unambiguous evidence for a differentially rotating disc seen edge-on supported by a number of independent facts. These include the symmetry and decrease in intensity from a central peak as well as the curvature of the $p, v$–diagram spine. Our modelling fits a ~ 100 point data matrix with only four free parameters to an overall accuracy of better than 3-4%. This is the highest ratio of observables to fitted parameters that we are aware of for any dynamical study claiming discs around a high-mass protostar.

Based on our present maser data our modelling gives estimated central masses for Keplerian discs is in the range $4\,M_\odot$ to infinity (see §5.4). As noted in §5.5 other observations of outflows in radio continuum and millimeter lines and of the total bolometric luminosity suggest a mass $>8\,M_\odot$ with the best estimate being $30\,M_\odot$. It seems clear that our differentially rotating gas disc surrounds a high-mass star but it is not yet clear what the mass of the central object is. Confirmation of a mass $> 20\,M_\odot$ would be highly significant because it would be beyond the mass threshold of the highest mass stars so far produced via accretion discs in hydrodynamic simulations (see e.g. Cesaroni et al. 2007; Zinnecker & Yorke 2007).

It is difficult to accurately constrain the central mass with our present maser data because we only detect emission from a small part of the disc (see §5.4). However prospects are good for making mass estimates with future deeper observations because for different central masses we can predict the full $p, v$–diagram and compare with these new observations. Figure 9 shows the full $p, v$–diagram distribution for Keplerian rotation around a central mass of $31\,M_\odot$; interestingly this already suggests some agreement with the data which was not part of the original fitting. In the figure the red contours correspond to the extent of the fitted data, the green dots indicate the position of the outliers at displacements around 40 and 80 mas (see Fig. 1). Even though these outliers were not considered in making the fit, the spine of the model passes close to at least one of them. This model also predicts local maxima in the maser emission at the edge of the disc (so called tangential maser emission) at displacements ±220 mas, at levels just below our present observational limits. Deeper observations can search for this emission. Such a detection would be a very powerful confirmation of the validity of our model and would more tightly constrain the central mass estimate. Another approach to constraining the mass is to look for the proper motions of distinct components relative to the central peak. In fact, in the framework of an edge-on rotating disc, if the outliers are indeed independent inhomogeneities orbiting the central star in a Keplerian fashion then they should show a transverse motion either toward or away from the central LOS, marked by the brightest emission. In particular, if the outlier closest to the main feature is at a radius of some 600 AU from a $30\,M_\odot$ central protostar we expect for Keplerian dynamics a proper motion of some 0.2-0.3 mas yr$^{-1}$, which should be detectable within the multiple epoch data that are presently being analysed (Jerkstrand et al. in prep.).

Finally, further investigations could cast light on the real shape of the $\eta$-profile. As stated in § 3 the shape of the $\eta$ function contains all details of inversion, temperature and density gradients, that we do not aim to discuss in this paper. Hints for tackling the exact modelling of the maser pumping come from the fact that 6.7 and 12.2 GHz maser maps are strikingly similar (see Fig. 1). This is in fact slightly surprising as the two masers are produced by two different molecules that in turn are expected to be excited differently and hence trace different regions. High resolution (millimetre interferometric) observations of thermally emitting transitions of methanol sharing levels with maser transitions will allow to directly estimate the level of amplifcation of both masers and possibily the level of inversion. These projects are ongoing and will be reported on in a later publication.

### 7.2. Summary and conclusion

In this paper we have carried out a detailed analysis of the main methanol maser feature in NGC 7538 IRS 1 N. We have extended upon the formalism and edge-on disc model first presented by Pestalozzi et al. (2004). In addition we have considered alternative outflow models. Our main conclusions are:

- the main feature of the 12.2 GHz (and 6.7 GHz) methanol maser emission toward NGC 7538 IRS 1 N is well modelled by a sufficiently fast differentially rotating disc seen edge-on;
- in contrast outflow models are unable to explain the central peak seen in the $p, v$–diagram and the gradual fall-off in opacity versus projected distance. They also do not naturally explain the curvature seen in the $p,v$-spine data. Only by resorting to extremely fine-tuned models can outflow models duplicate the observed features in the data;
- for Keplerian rotating discs we find good fits for central masses ranging from $4\,M_\odot$ to infinity. Models with central masses of of ~$30\,M_\odot$, as required by the bolometric luminosity, provide excellent fits;
- because of the high ratio of observables to unknowns in our modelling we argue that the NGC 7538 IRS 1 N maser emission provides one of the best pieces of evidence to date for a compact accretion disc around a high-mass protostar. Our results further bolster the case for massive stars forming via the same accretion disc mechanism at work in low-mass stars;
- for different central masses our disc model makes clear predictions for the opacity in the full $p, v$–diagram and proper motions. These predictions can be tested against future observations to more accurately constrain the central mass.

Although this study provides strong support for a compact disc structure for the main methanol maser feature in NGC 7538 IRS 1N, it seems certain that not all class II methanol masers occur in discs. The interpretation of ordered lines of methanol masers in terms of disc or outflow models must be decided case by case with detailed modelling. Even in NGC 7538 itself, the weaker methanol masers to the south, which are associated with other maser species (OH, $H_2O$ etc), will have to be investigated separately to decide whether they arise from an outflow or a disc.

An important final point is that the application of our formalism is not restricted to masers but can also be applied to optically thin thermal spectral line emission. For such emission the brightness is directly proportional to the optical depth ($I_{thin} \propto \tau$), while for masers $I_{maser} \propto e^\tau$. This paper has investigated model $\tau$ distributions from edge-on rotating discs, which were compared to the natural logarithm of the brightness of our observed maser emission. For optically thin thermal edge-on discs[5] the same disc model $\tau$ can instead be compared directly to the intensity. One subtle point should be noticed however: the average over a varying exponential function in $p,v$-space is not the same as taking the average of opacity and exponentiating it. This issue has to be taken into account if we compare models and data in maser systems which we do not resolve in space or velocity, or when we consider calculating integrated spectra of thermal and maser sources. In the latter case, for the same disc geometry, kinematics and $\eta(\rho)$ the total integrated spectra arising from maser emission would *not* in general equal the exponential of the spectrum from optically thin thermal emission. With the above practical caveats the results from this paper can be applied to (nearly) edge-on thermal discs. These results include the equations for opacity in the $p, v$–diagram (eq. 3) and opacity spine (eq. 6), the inner gradient of this spine (eq. 7) and its intensity variation (eq. 8) as well as its curvature (eq. 10). The condition for having a central peak in opacity (eq. 9) also applies.

*Acknowledgements.* This work is the result of a fruitful collaboration involving people from several institutes in Europe, USA and Chile aiming at studying the high-mass star forming region NGC 7538 (see also http://physics.gu.se/~micpes). The fitted 12 GHz data that has fitted was reduced in collaboration with A. Jerkstrand. Michele Pestalozzi thanks Gothenburg University for support through a post-doctoral position. Moshe Elitzur acknowledges NSF support through grants AST-0507421 and AST-0807417. John Conway acknowledges partial support by a Swedish Research Council grant.

---

[5] For non-edge-on optically thin thermal discs the formalism of this paper can be applied to 2D position-velocity data created from the full data cube by integrating emission along the minor axis of the projected disc